Scalable production of high sensitivity, label-free DNA biosensors based on back-gated graphene field-effect transistors

Jinglei Ping[1,†], Ramya Vishnubhotla[1,†], Amey Vrudhula[2] & A. T. Charlie Johnson[1,*]

[1]Department of Physics and Astronomy, University of Pennsylvania, Philadelphia 19104, United States

[2]Department of Electrical and Systems Engineering, University of Pennsylvania, Philadelphia 19104, United States

[†]These authors contributed equally to this work.

**Abstract:** Scalable production of all-electronic DNA biosensors with high sensitivity and selectivity is a critical enabling step for research and applications associated with detection of DNA hybridization. We have developed a scalable and very reproducible (> 90% yield) fabrication process for label-free DNA biosensors based upon graphene field effect transistors (GFETs) functionalized with single-stranded probe DNA. The shift of the GFET sensor Dirac point voltage varied systematically with the concentration of target DNA. The biosensors demonstrated a broad analytical range and limit of detection of 1 fM for 60-mer DNA oligonucleotide. In control experiments with mismatched DNA oligomers, the impact of the mismatch position on the DNA hybridization strength was confirmed. This class of highly sensitive DNA biosensors offers the prospect of detection of DNA hybridization and sequencing in a rapid, inexpensive, and accurate way.

All-electronic DNA biosensors offer considerable promise for rapid genetic screening and nucleic acid detection for gene-expression investigations, pharmacogenomics, drug discovery, and molecular diagnostics[1]. In order to enable these applications, the electronic DNA biosensors need to be sensitive, selective, and based upon a scalable fabrication process. Wafer-scale graphene, a one-atom thick sheet of carbon with remarkable electronic sensitivity, outstanding biocompatibility[2], and extremely low signal-to-noise ratio[3], can be prepared via chemical vapor deposition[4,5]. However, very few previous reports on graphene field-effect-transistors (GFETs) for DNA sensing[6-8] were based on scalable fabrication methods. To this point there are no reports of more than 10 functional devices fabricated on a single chip, and the sensitivity has been limited to 100 fM[7].

Here we describe the development of scalable DNA biosensors based on back-gated GFETs with sensitivity as low as 1 fM (~6×10$^5$ DNA molecules in a 1 mL drop). We prepared graphene by chemical vapor deposition (CVD) and fabricated GFETs with conventional photolithography. The GFETs demonstrated high yield (> 90%) and consistent transport properties. The GFETs were functionalized using a well-controlled chemical treatment that enabled high surface coverage with single-stranded probe DNA. DNA biosensors created in this way exhibit a wide analytical range (three decades in concentration) and excellent selectivity against non-complementary DNA oligomers. The sensitivity of the DNA biosensors depends systematically on the length of the oligomer, and for 60-mer DNA 1 fM limit of detection was achieved, 100-fold improvement over earlier reports. The response calibration curves of the DNA biosensors were in excellent agreement with predictions of the Sips model[9] for DNA-hybridization. Our control experiments confirmed that sensor responses were determined by hybridization between the probe and target DNA oligomers, and the results were consistent with earlier reports of hybridization using

DNA microarrays. Our methodology has the potential to be developed into a rapid and convenient point-of-care tool with clinically relevant sensitivity.

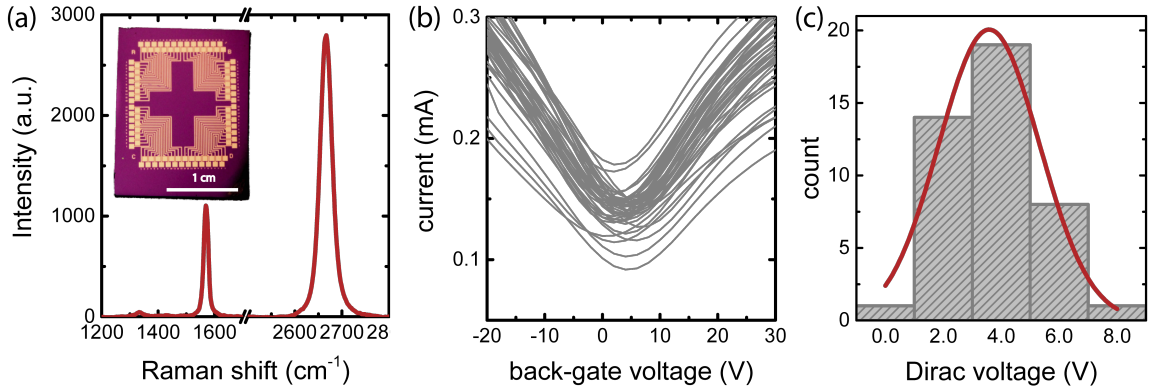

Figure 1. (a) Raman spectrum of the channel region of a graphene field effect transistor (GFET) after processing. Inset: Optical micrograph of an array of 52 GFETs. (b) $I$-$V_g$ characteristics for an array of 52 GFET devices showing excellent reproducibility. (c) Histogram of the Dirac voltage extracted from the $I$-$V_g$ characteristics of panel (b) along with a Gaussian fit to the data (red curve).

A 2.5 x 2.5 cm graphene sample was prepared via chemical vapor deposition on a copper growth substrate and transferred using an electrolysis bubbling method[5, 10] onto a 2 x 2.5 cm oxidized silicon substrate with pre-fabricated, 45-nm thick Cr/Au electrodes for an array of 52 GFETs. We find that this transfer method effectively limits contamination, doping, and damage associated with graphene transfer. The GFET channels were then defined using photolithography and oxygen plasma etching (Fig. 1a, inset). The sensor array was cleaned by annealing in an argon/hydrogen atmosphere before further characterization or chemical functionalization. (See Methods for additional details of the fabrication process) This method is compatible with scale up to thousands of GFETs or more, as well as integration with prefabricated CMOS signal processing circuitry[11].

The graphene in the GFET channels was single-layer with low defect density, as verified by the 2D/G ratio (~2) and the minimal D peak intensity in the Raman spectrum[12] (Fig. 1(a)). The excellent quality of the graphene enables consistent GFET transport properties and high fabrication yield (>90%), based on more than 30 arrays fabricated for this experiment. As shown in Fig. 1(b), the current-gate voltage ($I$-$V_g$) characteristics for all 52 GFETs in a single array are very similar. The Dirac point of the GFETs, where the $I$-$V_g$ characteristic has a minimum, lies in a narrow range near zero back-gate voltage, 3.6 ± 4.0V (Fig. 1(c)), indicating low doping effects induced in our methodology.

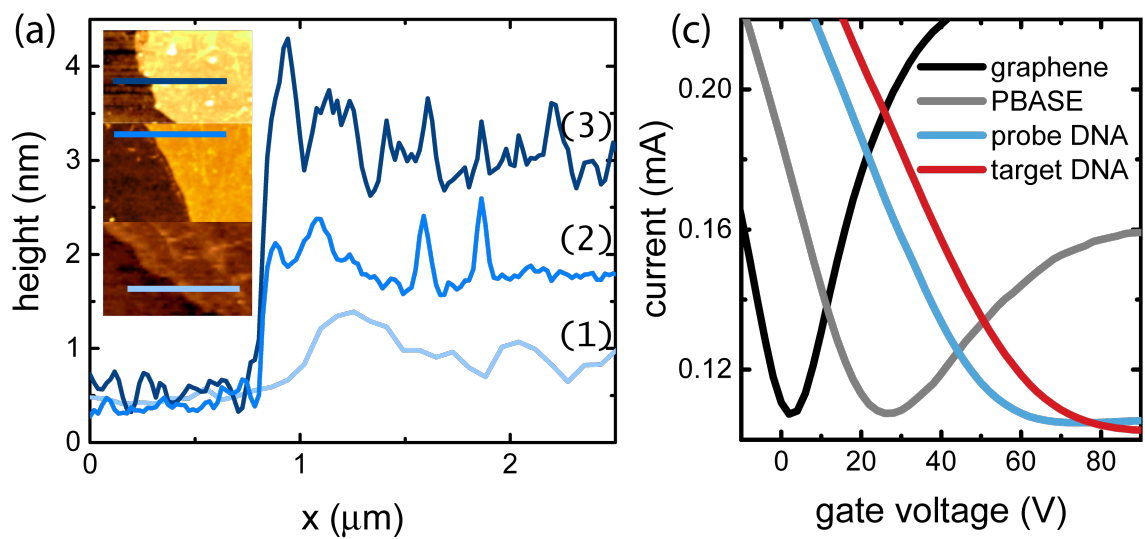

Figure 2. (a) AFM line scans of (1) annealed graphene, (2) PBASE-functionalized graphene, and (3) graphene functionalized with PBASE and aminated DNA. Inset: AFM images showing the scan lines plotted in the main figure. Scan lines are 2.5 μm. Z-scale 8 μm. (b) $I$-$V_g$ characteristics for a typical GFET that was annealed, functionalized with PBASE, reacted with 22mer aminated probe DNA, and exposed to 10 nM target DNA in deionized water.

After annealing, the GFET channels were functionalized by incubation for 20 hrs in a solution of the bifunctional linker molecule 1-Pyrenebutyric acid N-hydroxysuccinimide ester (PBASE) in dimethylformamide (DMF) (See Methods for details). The aromatic pyrenyl group of PBASE binds to the basal plane of graphene through the non-covalent π-π interaction[13, 14]. This process yields a uniform, ~ 1 nm thick monolayer[15] of self-assembled PBASE on the graphene (see linescan (1) in Fig. 2(a)), except at wrinkles (~ nm high) in the CVD graphene created by the transfer process[16]. The aminated (5') probe DNA (22mer, 40mer, or 60mer) was then bound to the PBASE linker by a N-hydroxysuccinimide (NHS) crosslinking reaction (See Methods for details). Due to the high coverage of the PBASE monolayer, the probe DNA molecules were immobilized on the graphene channel at such high density that individual DNA molecules could not be distinguished in AFM images acquired using a conventional AFM cantilever, with tip radius ~ 100 nm (Fig. 2(a)). The average height increase of the GFET due to attachment of the 22mer probe DNA is ~ 1.2 nm, consistent with the molecular size. After attachment of the probe DNA, GFET DNA biosensors were tested against the complementary single strand DNA "target" and various controls.

The $I$-$V_g$ characteristics of the GFET devices were measured in the dry state[5] after each step of functionalization chemistry and again after exposure to the target. The value of the Dirac voltage for each $I$-$V_g$ characteristic is determined using a curve-fitting method[17] through the equation

$$I^{-1} = \left[e\alpha\mu(V_{bg} - V_D)\right]^{-1} + I_s^{-1}.$$

Here $I$ is the current, $\mu$ the mobility, $V_{bg}$ the back-gate voltage, $V_D$ the Dirac voltage, $\alpha$=7.2 · $10^{16}$ cm$^{-2}$ V$^{-2}$ the constant relating gate voltage to carrier number density, and $I_s$ the saturation current due to short-range scattering[18]. Formation of the PBASE monolayer leads to an increase in $V_D$ of ~ 23 ± 3.3 V (Fig. 2(b)). This is explained by considering chemical gating effects associated with residual water on the device surface. Here, we assume that NHS groups are hydrolyzed into carboxyl groups, which deprotonate and acquire a negative charge. Attachment of 22mer probe DNA led to a further 40 V increase in the Dirac voltage, which is explained quantitatively through the chemical-gating effect[17] of probe-DNA molecules that become negatively charged due to ionization of phosphate groups in residual water. This Dirac voltage shift corresponds to an increase in the positive (hole) carrier density in the graphene by ~ 3.0 x $10^{12}$ cm$^{-2}$. Assuming chemical gating of 22 negative charges for each oligomer, the density of immobilized probe DNA is ~ 1.3×10$^3$ µm$^{-2}$, more than an order of magnitude higher than the level of protein attachment achieved using a very similar functionalization approach [5, 17]. This corresponds to a separation of ~ 25 nm between DNA molecules, a factor of 4 lower than the tip radius and so consistent with the uniform DNA coverage observed by AFM (Fig. 2(a)).

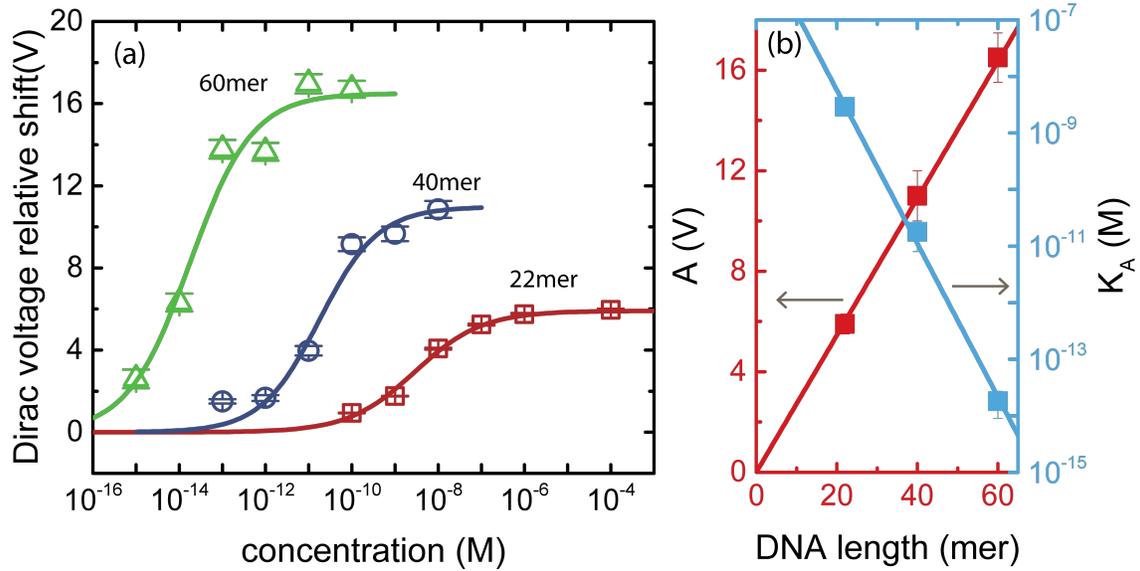

Figure 3. (a) Relative Dirac voltage shift as a function of concentration for DNA targets of different lengths. Error bars (standard deviation of the mean) are approximately equal to the size of the plotted point. Solid curves are fits to the data based on the Sips model. (b) Variation of the fit parameters $A$ (red data) and $K_A$ (blue data) in Eqn. (1) with DNA oligomer length. The red and blue lines are fits to the data, as discussed in the main text.

In response experiments, all 52 GFET sensors on a single chip were tested against a solution with a known concentration of target DNA or a related control in deionized water. The Dirac voltage of the $I$-$V_g$ characteristic showed a reproducible shift to positive voltage, $\Delta V_D$, as seen in Fig. 2(c). To compare results across the three different DNA targets, for each concentration tested we plot the Dirac voltage shift relative to $\Delta V_D^0$, the shift measured upon exposure to pure deionized water, i.e., $\Delta V_D^{REL} = \Delta V_D - \Delta V_D^0$, with the results shown in Fig. 3(a). For all DNA oligomers tested, the relative shift varied systematically with target concentration, and it is ascribed to an increase in the positive carrier concentration in the GFET channel induced by the negatively-charged phosphate groups of target DNA molecules that have hybridized with probe DNA on the GFET surface.

The Sips model[9, 19] for describing DNA hybridization provides an excellent fit to the measured data data for $\Delta V_D^{REL}$ as a function of target concentration:

$$\Delta V_D^{REL} = A \frac{(c/K_A)^a}{1+(c/K_A)^a} \quad (1)$$

where $c$ is the concentration of the target DNA solution, $A$ the maximum response with all binding sites occupied, and $K_A$ the equilibrium dissociation constant. The parameter $a$ in the Sips model represents a Gaussian distribution of DNA binding energies where $a=1$ corresponds to single binding energy level. The best fit to the data for the 22mer target yields fit parameter values: $A$ = 5.9 ± 0.4 V, $K_A$ = 2.9 ± 0.9 nM, and $a$ = 0.56 ± 0.07. The analytic range of the fit (Fig. 3(a)) covers three orders of magnitude, from ~100 pM to ~100 nM, with sensitivity ~1.4 V/decade. The dataset presented in Fig. 3(a) indicates that GFET-based DNA biosensors can differentiate between DI water and a solution containing the 22mer target at a concentration of < 100 pM. The best fit value of $K_A$, 2.4 ± 0.8 nM, agrees well with that expected for 20-mer DNA hybridization[20], 1.7 nM. The best fit value of $a$ = 0.56 ± 0.07 implies a heterogeneous adsorption isotherm with a distribution of binding energies[9, 19] rather than a single value DNA-DNA binding energy, which would yield $a$ = 1. This binding-energy distribution is assumed to reflect significant interactions between the probe and/or target DNA and the graphene surface[21].

|  | $A$ | $K_A$ | $a$ |
| --- | --- | --- | --- |
| 22mer | 5.9 ± 0.4 V | 2.9 ± 0.9 nM | 0.56 ± 0.07 |
| 40mer | 11.0 ± 1.0 V | 17.8 ± 9.8 pM | 0.64 ± 0.14 |
| 60mer | 16.5 ± 1.0 V | 18.1 ± 9.0 fM | 0.60 ± 0.17 |

Table 1. Fitting parameters for all probe DNA sequences tested.

We also tested GFET DNA biosensors based on 40mer probe DNA and 60mer probe DNA. As shown in Fig. 3(a), the limit of detection (LOD) using 40mer probe DNA is ~ 100 fM, and the 60mer target DNA was reliably detected at a concentration of 1 fM (~6×10$^5$ DNA molecules in a 1 mL drop), to our knowledge, the most sensitive DNA detection reported to date. The Sips model fit parameters for the three probe DNA sequences are shown in Table 1. The distribution function index is roughly the same for the different DNA targets, indicating comparable degree of binding energy heterogeneity. The fit values for $A$ and $K_A$ demonstrate two advantages of using longer DNA oligomers. First, the maximum signal level (A) increases nearly linearly with DNA length, at a rate of 0.27 V/mer (see Fig. 3b, red data points). This is assumed to reflect that the charge carried by each DNA chain increases as the DNA length increases, enhancing the chemical gating effect on the graphene and leading to a proportionately larger Dirac voltage shift. Second, the dissociation constant decreases exponentially for longer DNA. As seen in Fig. 3(b), in the log($K_A$)-length relationship is approximately linear, with slope of -0.225±0.024. This is in good agreement with the slope of 0.138±0.006 that was found using a quartz crystal microbalance approach[20].

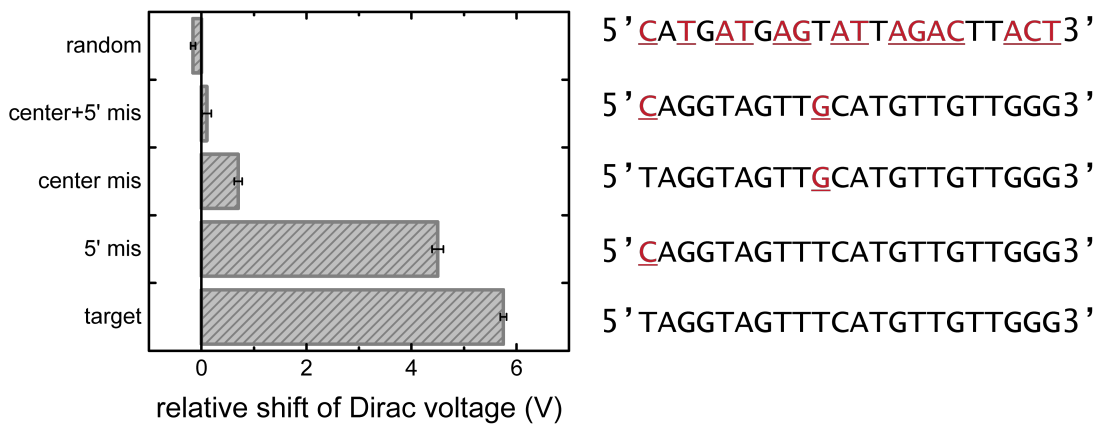

Figure 4. Relative response of GFET-based 22mer DNA biosensors to the target sequence and various controls, all at a concentration of 1 µM. The base sequences of the oligomers tested are listed, with mismatches shown in red. Starting from the bottom, the oligomers tested are: target DNA, single mismatch at the 5' end, single mismatch at the center, two mismatches at the 5' end and the center, and random sequence DNA. Error bars are standard deviation of the mean.

Multiple control experiments were conducted with 22mer DNA biosensors to verify that the biosensors responses reflected specific binding of the complementary target DNA. A variety of control samples were used, with all control solutions having a concentration of 1 µM in DI water. In Fig. 4, we report the results as the Dirac voltage shift induced by the target or control at a concentration of 1 µM, relative to the shift induced by pure DI water. The target 22mer DNA gives the largest value for the relative Dirac Voltage shift ($\Delta V_D^{REL}$=5.7 ± 0.4 V), which is expected since it should have the highest binding affinity for the probe DNA and therefore the largest associated change in GFET carrier concentration. The single base-mismatch controls are expected to interact more weakly with the probe DNA. It is intriguing to note that the control with a single base-mismatch at the 5' end shows a slight response decrease (~ +4.6 ± 0.7 V or 80 ± 12% of that for the target DNA), while the response to control DNA with the mismatch at the center is strongly suppressed (+0.7 ± 0.5 V), only ~10% of that for the target DNA. Experiments based on DNA oligonucleotide microarrays[22, 23] show similar effects in how the response depends on the position of a single base mismatch. The reason why a mismatch at the center of the strand has such a strong effect on hybridization can be understood through a positional-dependent-nearest-neighbor model[23, 24]. The control oligomer with two mismatches, one at the center and one at the 5' end, gave a sensor response that was indistinguishable from the response to DI water, and the same was true for the response to a 1 μM solution of a random sequence DNA oligomer (32% consistent with the target DNA).

In conclusion, we have developed a scalable fabrication approach for arrays of graphene-based DNA biosensors with all-electronic readout, and we measured their responses to the complementary DNA target and multiple control oligomers. The fabrication process is based upon conventional photolithographic processing and should be suitable for mass production. The GFETs fabricated for the experiments were of very high quality, as evidenced by Raman spectroscopy, Atomic Force Microscopy, and electronic measurements. The DNA biosensors have a wide analytical range and a sensitivity that depends systematically on the length of the DNA. For 60mer DNA, we achieved a detection limit of 1 fM, 100x lower than previous reports for graphene-based DNA biosensors. Measured sensor responses over a range of six orders of magnitude in concentration were well fit by the Sips model. Control experiments verified that the sensor response was derived from specific binding of the probe DNA to the target DNA, and also confirmed that the complementary DNA with a mismatch at the center hybridizes much more weakly with the probe DNA than at the 5' end. This advance in graphene-enabled biosensor technology paves the way for all-electronic and cost-effective DNA biosensors for quantitative investigations and clinical applications.

**Methods.** *CVD growth of large-area graphene.* Monolayer graphene was grown via low pressure chemical vapor deposition on a copper foil substrate (99.8%, 25 μm thick, Alfa Aesar) in a four-inch quartz tube furnace. The copper was annealed for 60 minutes at 1020 °C in ultra-high purity hydrogen (99.999%; flow rate 80 sccm). Graphene was then synthesized using methane as a precursor (temperature 1020 °C; hydrogen flow rate 80 sccm; methane flow rate 10 sccm; pressure 850 mT; growth time 20 min). The tube was then cooled to room temperature in 40 minutes.

*Graphene Transfer.* A sacrificial layer of 500-nm thick poly(methyl methacrylate) (PMMA) was spin-coated on top of the graphene/copper substrate. The sample was baked for 2 minutes at 100 °C, then connected to the cathode of a power supply and immersed in a 50 mM sodium hydroxide solution in DI water. A constant voltage of 20 V was applied between the cathode and a platinum anode also in the electrolyte solution. Under these conditions, hydrogen bubbles were generated between graphene and the copper foil, causing the PMMA/graphene film to detach from the copper substrate. The PMMA/graphene film was then washed in a series of DI-water baths and transferred onto the surface of a 300 nm $SiO_2$/Si substrate on which Cr/Au electrodes had previously been fabricated. The sample was left to dry for one hour, then baked at 150 °C for 2 minutes before the PMMA film was removed by washing with acetone and isopropyl alcohol. The sample was then dried with compressed nitrogen.

*GFET array fabrication.* A protective layer of Polymethylglutarimide (PMGI, Microchem) was spin-coated on the graphene and baked on a hot plate at 125 °C for 5 minutes. Next, a layer of S1813 (Microchem) was spin coated atop the PMGI and the chip was baked on a hot plate at 100 °C for 2 minutes. The channel regions of the GFETs were defined using optical lithography, and the chip was developed in MF-319 (Microposit) according to the manufacturer's instructions, cleaned with DI water, and finally dried using compressed nitrogen gas. Graphene outside of the defined FET channels was removed by oxygen plasma etching for 30 seconds (power 80 W, O2 gas pressure 1.25 Torr). Photoresist residue on the chip was removed by soaking in acetone (5 minutes), photoresist remover 1165 (Microposit, 5 minutes), and acetone

(30 minutes) followed by a final rinse with isopropyl alcohol. The GFET array was then annealed in tube furnace under flowing hydrogen (250 sccm) and argon (1000 sccm) at 225 °C for 1 hour.

*PBASE-functionalization, probe-DNA-immobilization, and testing against target or control solution.* The GFET array was soaked in a 1 mM PBASE (Sigma-Aldrich) in DMF (Fisher) solution for 20 hours and then washed thoroughly with DMF, IPA, and DI-water, each for three minutes. The array was then incubated in 1 μM aqueous solution of 22mer probe-DNA (5'-Amine-CCCAACAACATGAAACTACCTA-3'), 40mer probe-DNA (5'-Amine-AATTACAAAAACAAATTACAAAAATTCAAAATTTTCGGGT-3') or 60mer probe-DNA (5'-Amine-AAACTAAAGAATTACAAAAACAAATTACAAAAATTCAAAATTTTCGGGTTTATTACAGGG-3') for three hours and rinsed with DI water. Following probe attachment, an aqueous solution of target or control DNA with known concentration was pipetted onto the chip, and incubated in a warm, humid environment for 30 minutes to suppress evaporation. The chip was rinsed with DI-water and dried with compressed nitrogen before performing electrical measurements.